\documentclass[12pt]{article}
\usepackage{microtype} \DisableLigatures{encoding = *, family = * }
\usepackage{url}
\usepackage{amsmath, amsthm,epsfig,bm, booktabs,amssymb}
\usepackage{amsmath, amsfonts}
\usepackage[listformat=empty]{caption}
\usepackage{amsmath,latexsym,amssymb,amsfonts,amsthm}
\usepackage[margin=3.5cm]{geometry}
\usepackage{color}
\usepackage{mathrsfs}
\usepackage{algorithm,algorithmic}
\usepackage{epstopdf}
\usepackage{threeparttable}
\usepackage{graphicx}
\usepackage{subfig}
\usepackage{tabularx}
\usepackage[all]{xy}
\usepackage{float}
\usepackage{cases}
\usepackage[bookmarksnumbered=true, pdfauthor={Zhe Sun}]{hyperref}

\newcommand{\be}{\begin{equation}}
\newcommand{\ee}{\end{equation}}
\newcommand{\ba}{\begin{align*}}
\newcommand{\ea}{\end{align*}}
\newcommand{\bea}{\begin{eqnarray*}}
\newcommand{\eea}{\end{eqnarray*}}
\newcommand{\bi}{\begin{itemize}}
\newcommand{\ei}{\end{itemize}}

\newcommand{\hf} {\hat{f}}

\newcommand{\baf} {f^*}
\newcommand{\bx} {x^*}
\newtheorem{example}{Example}[section]
\newtheorem{thm}{Theorem}[section]
\newtheorem{defin}{Definition}[section]
\newtheorem{assump}{Assumption}[section]

\newtheorem{rmk}{Remark}[section]

\newcommand{\rank}{\mathrm{rank}}

\newcommand{\Ker}{{\rm Ker}}
\newcommand{\Ran}{{\rm Ran}}

\newcommand{\Rl}{\mathbb{R}^l}
\newcommand{\Rm}{\mathbb{R}^m}
\newcommand{\R}{\mathbb{R}}

\newcommand{\Rnl}{\mathbb{R}^{n\times l}}
\newcommand{\Rmn}{\mathbb{R}^{m\times n}}
\newcommand{\Rnn}{\mathbb{R}^{n\times n}}
\newcommand{\Rn}{\mathbb{R}^{n}}
\newcommand{\0}{\mathbf{0}}
\newcommand{\M}{\mathcal{M}}
\newcommand{\K}{\mathcal{K}}
\newcommand{\Ss}{\mathcal{S}}
\newcommand{\I}{\mathcal{I}}
\newcommand{\T}{\top}

\newcommand{\comments}[1]{}

\title{ $\ell_1$-minimization method for link flow correction} 
\author{Penghang Yin\thanks{Department of Mathematics, University of California, Los Angeles, Los Angeles, CA, 90095. Email: yph@ucla.edu.}
, Zhe Sun\thanks{Department of Civil and Environmental Engineering, Institute of Transportation Studies, 4040 Anteater Instruction and Research Bldg, University of California, Irvine, CA 92697-3600. Email: zhes@uci.edu}, 
Wen-Long Jin\thanks{Department of Civil and Environmental Engineering, California Institute for Telecommunications and Information Technology, Institute of Transportation Studies, 4038 Anteater Instruction and Research Bldg, University of California, Irvine, CA 92697-3600. Email: wjin@uci.edu. Corresponding author.},
and 
Jack Xin\thanks{Department of Mathematics, University of California, Irvine, Irvine, CA 92697. Email: jxin@math.uci.edu. }
}
\date{}
\begin{document}
\maketitle

\begin{abstract}
A computational method, based on $\ell_1$-minimization, is proposed for the problem of link flow correction, when the available traffic flow data on many links in a road network are inconsistent with respect to the flow conservation law. Without extra information, the problem is generally ill-posed when a large portion of the link sensors are unhealthy. It is possible, however, to correct the corrupted link flows \textit{accurately} with the proposed method under a recoverability condition if there are only a few bad sensors which are located at certain links.  We analytically identify the links that are robust to miscounts and relate them to the geometric structure of the traffic network by introducing the recoverability concept and an algorithm for computing it. The recoverability condition for corrupted links is simply the associated recoverability being greater than 1. In a more realistic setting, besides the unhealthy link sensors, small measurement noises may be present at the other sensors. Under the same recoverability condition, our method guarantees to give an estimated traffic flow fairly close to the ground-truth data and leads to a bound for the correction error. Both synthetic and real-world examples are provided to demonstrate the effectiveness of the proposed method.
\end{abstract}


\section{Introduction}

Link volume/flow data is an important data source in both long-term planning and short-term operation applications.  The examples include but are not limited to signal timing, toll road pricing, origin-destination trip matrix estimation, transportation planning, traffic safety (e.g. \cite{STM_08, Lindsey_06, McNally_07, Lord_10} and the references therein).

The flow conservation in a traffic network implies that the total in-flow equals the total out-flow at each non-centroid node. The centroids are nodes where traffic originates/is destined to, and non-centroids nodes denotes all the other nodes. Practically, when looking at traffic flow counts over a sufficiently long time period (e.g. daily cumulative flow), we expect that the sum of cumulative link flows entering the non-centroid node equals the sum of cumulative link flows leaving it.  

The flow conservation law is an important property, which has been exploited in many different applications. For example, the widely used first-order traffic flow model, the LWR model \cite{LW_55,Richards_56}, is derived based on the conservation of traffic. In \cite{Chen_09}, the authors mentioned that a path flow estimator (PFE) needs reasonably consistent link flows, meaning that the flow conservation law should be satisfied within a certain error bound, to reproduce feasible path flow solutions. 

In practice, the flow conservation law can be violated due to numerous flow measuring errors; i.e., the observed flow counts are generally corrupted and cause data inconsistency issues. In \cite{Sun_16},  the network sensor health problem (NSHP) (\cite{Sun_16}) is proposed to evaluate  individual sensors' health indices based on the level of flow data consistency. Assuming flow counting sensors are already installed on some of the links where at least one base set exists, the NSHP tries to find the least inconsistent base set that ``minimizes the sum of squares of the differences between observed and calculated link flows''. The health index of a specific sensor is evaluated based on the frequency that it appears in the least inconsistent set. 

Several studies have looked into the problem of correcting inconsistent flow data according to flow conservation. To solve a similar problem in transit planning, Kikuchi et al. \cite{Kik_71} studied the passenger flow  balancing problem and proposed a least square correction method to adjust the flows, so that the counts are conserved and close to the observed values. van Zuylen and Branston \cite{VanZuylen_82} assumed that the observed link flows follow probability distributions constrained by flow conservation. The study derived the formula for constrained maximum likelihood estimates of the link flows.  Kikuchi et al. \cite{Kik_00} examined and compared six different methods
to adjust observed flow rate according to flow conservation. All of the methods have the same constraints but different objective functions. Vanajakshi and Rilett \cite{Vana_04} studied flow inconsistency problem between neighboring upstream and downstream loop detectors. A nonlinear optimization problem is proposed to correct loop detector data, in the case when observed data violates flow conservation. 

In summary, given the observed cumulative flows on different links, all of the existing flow correction methods adopted optimization approaches that try to meet the following principles:
\begin{itemize}
	\item Ensure that flow conservation be followed exactly at all non-centroid nodes after adjustment using a set of constraints,
	\item Preserve the integrity of the observed data as much as possible by minimizing the distance between adjusted and observed flows.
\end{itemize}
However, all of the studies are limited to simple hypothetical networks or networks with simple topologies. Also, no systematic study has been done regarding the effectiveness and applicability of the methods.

In this study, we propose a method to estimate the true link flow from corrupted data on 
observed links as well as unobserved links via $\ell_1$-minimization. 
Similar to the existing methods, the link flow correction method is also 
formulated as an optimization problem to minimize the difference between observed and 
estimated link flows. As an improvement over the existing methods, 
the node-based formulation of flow conservation is introduced to handle 
general road network where link flows are only observed on monitored links, 
not on all links as assumed in many existing studies. More importantly, 
we adopt the $\ell_1$-minimization method from compressed sensing \cite{CRTV_05,CJT_06} to analytically derive 
the condition for exact/stable recovery of the true cumulative flow counts.The $\ell_1$ norm is 
the unique convex sparsity promoting penalty. Though it is not differentiable, 
various efficient scalable numerical methods exist to date for its minimization \cite{BT_09,Boyd_11,Daub_04,GO_09,YZ_11} 
besides linear programming.
In addition to $\ell_1$ norm, other non-convex sparsity promoting penalty 
functions can also be considered; see \cite{YLHX,NonCvxL1,LYX} and references therein.  
Their minimization is computationally more expensive than $\ell_1$, and
we shall leave such a study for a future work.

The rest of the paper is organized as follows. In section \ref{sec:method}, we state the link flow 
correction problem formulation, the exact and stable recovery theorem, the recoverability condition and 
the connection with compressed sensing. In section \ref{sec:estimation}, we use a toy example to illustrate the conditions for exact and stable link flow recovery. In section \ref{sec:real-world}, we use real-world loop detector data as an 
application for this method. In both the toy and real world examples, the recoverability condition is verified 
analytically. The concluding remarks are in section \ref{sec:conclusion}.

\subsection*{Notations}
Let us fix some notations. $\R^n$ represents the real coordinate space of $n$ dimensions. Let $x\in\Rn$, $\|x\|_1:= \sum_{i=1}^{n}|x_i|$ takes the $\ell_1$ norm of $x$, and $\|x\|$ denotes the Euclidean ($\ell_2$) norm. Given any index set $\I\subseteq\{1,2,\dots,n\}$, $|\I|$ counts the number of elements in $\I$; $\I^c := \{1,2,\dots,n\}\setminus\I$ is the complement set of $\I$. $x_\I\in\R^{|\I|}$ consists of the elements in $x$ restricted to the index set $\I$. $\0_{(n)}\in\Rn$ denotes the vector containing zeros only, while $I_{(n)}\in\Rnn$ denotes the identity matrix of order $n$. For any matrix $A\in\Rmn$, $A^\T$ is the transpose of $A$; $A_\I\in\R^{|\I|\times n}$ is the submatrix of $A$ restricted to the row index set $\I\subseteq\{1,2,\dots,m\}$, and $A^\I\in\R^{m\times |\I|}$ is the submatrix of $A$ restricted to the column index set $\I\subseteq\{1,2,\dots,n\}$; e.g., $A_{\{1,2\}}$ extracts the first two rows of $A$, and $A^{\{1,2\}}$ extracts the first two columns of $A$. $\Ker(A):= \{x\in\Rn: Ax = \0_{(m)}\}$ represents the kernel space of $A$, while $\Ran(A):= \{h\in\Rm: h = Ax \mbox{ for some } x\in\Rn\}$ represents the range space of $A$.

\section{Methodology}\label{sec:method}
\subsection{Problem setup}
Given a traffic network with non-centroid nodes only, the node-link incidence matrix $A\in\Rnl$ with $n$ being the number of nodes and $l$ the number of links, can be expressed as
\begin{equation*}
A_{ij} = 
\begin{cases}
-1 & \mbox{if the $j$-th link is outgoing link of node i}\\
1 & \mbox{if the $j$-th link is incoming link of node i}\\
0 & \mbox{otherwise.}\\
\end{cases}
\end{equation*}
Then $A$ is always of full (row) rank as proved in \cite{Ng_12}, and
traffic flow data $\hf\in\Rl$ obeys the flow conservation:
\begin{equation}\label{eq1}
A\hf = \0_{(n)}.
\end{equation}
Suppose $\M\subseteq \{1,2,\dots,l\}$ is the set of links whose link flows are observed, and $|\M| = m$. We call $M$ as ``monitored set'' thereafter. We assume that 
\begin{displaymath}
f_{\M} = \hf_{\M} + e_{\M}\in\Rm,
\end{displaymath}
is the observed inconsistent flow data corrupted by sensing errors $e_{\M}\in\Rm$. 

\medskip

The flow correction problem is to derive an estimate of $\hf$, denoted by $f^*$, from the corrupted data $f_\M$. Here we impose an underlying assumption on $\M$ for the flow correction problem to be well-posed. We will need the concept of base set introduced in \cite{Sun_16}. 

\begin{assump}\label{ass}
$\M$ contains at least one base set $\K\subseteq \M$, meaning that $|\K| = l-n$ and $A^{\K^c}\in\Rnn$ is invertible.
\end{assump}

For the consistent data $\hf_\M = f_\M$ (i.e., $e_\M = \0_{(m)}$), of course we have $\hf_\K = f_\K$ since $\K\subseteq \M$. Then the $\hf$ can be uniquely recovered by performing \cite{Ng_12, Sun_16}:
$$
\hf_\K = f_\K \quad \mbox{and} \quad \hf_{\K^c} = -(A^{\K^c})^{-1}A^{\K}f_\K.
$$
If $\M$ contains more than one base set, the $\hf$ recovered in the above from different $f_\K$ will be consistent.

Assumption \ref{ass} is the sufficient and necessary condition for the whole link flows to be observable. It guarantees that the whole flow data can be deduced from at least one subset of the observed link flows. Without this assumption, however, some of the link flows cannot be estimated from available data and the problem is unsolvable \cite{Ng_12}, whether the measured flows are consistent or not. 

\subsection{Flow correction via $\ell_1$-minimization}
Since $A$ is of full row rank, $\Ker(A)$ is an $(l-n)$-dimensional subspace of $\Rl$. Suppose $Z\in\mathbb{R}^{l \times (l-n)}$ is the matrix whose columns form a basis of $\Ker(A)$. Since $\hf\in\Ker(A)$, we have
$$
\hf = Zx, \; \mbox{for some } x\in\R^{l-n}. 
$$  
As a result, $\hf_\M$ must be of the form $Z_\M x$ for some $x\in\R^{l-n}$. 

\begin{rmk}
Clearly the existence of $Z$ is non-unique, but $\baf$ is invariant to the choice of $Z$ and only depends on the structure of the traffic network. Indeed $f^*$ is the one in $\Ran(Z)$ whose restriction on $\M$ has the least absolute deviation from $f_\M$. So $\baf$ only depends on $\Ran(Z)$ which is same as $\Ker(A)$. Note that $A$ is the node-link matrix uniquely determined by the network structure.
\end{rmk}

The following result not only gives a concrete construction of $Z$, but also interprets $x^*$ in  (\ref{L1}) as an estimate of $\hf_{\K}$ for some base set $\K$ (not necessarily a subset of $\M$).

\begin{thm}\label{thm:ker}
Let $\K$ be any base set. Without loss of generality, suppose $A$ is partitioned as $[A^{\K}, \; A^{\K^c}]$ with $A^{\K^c}\in\Rnn$ being invertible. Then 
$$
Z = 
\begin{bmatrix}
I_{(l-n)} \\
-(A^{\K^c})^{-1}A^{\K}
\end{bmatrix}
\in\R^{l\times(l-n)}
$$
is a basis matrix of $\Ker(A)$. Moreover, by choosing such $Z$, $x^*$ from (\ref{L1}) is an estimate of $\hf_{\K}$. 
\end{thm}

We will show the proof in Appendix C. Our proposed method consists of the following two steps:
\bi
\item[1.] We first solve an $\ell_1$-minimization problem:
\begin{equation}\label{L1}
\bx = \arg\min_{x\in\R^{l-n}} \|Z_{\M}x - f_{\M}\|_1,
\end{equation}
That is, we seek an estimate of $e_\M=f_\M - \hf_\M$ in the affine space $\{f_\M - Z_\M x: x\in\R^{l-n}\}$ with the least $\ell_1$ norm. The problem (\ref{L1}) can be efficiently solved by the alternating direction method of multipliers (ADMM) \cite{Boyd_11}; see Appendix A for the implementation details. 

\item[2.] 
$\hf$ is then estimated by
\begin{equation}\label{Recon}
\baf = Z\bx.
\end{equation}
$Z\bx$ may have non-integer entries, in this case, we can just perform rounding.
\ei

\subsection{Connections with compressed sensing}
Compressed sensing \cite{CJT_06, Don_06} aims to recover a sparse signal (vector) $y$ from an under-determined linear system that generally has infinitely many solutions. It enables recovery of the signal $y$ from far fewer samples than required by the Nyquist-Shannon sampling theorem. Major ingredients of the standard compressed sensing technique include
\bi
\item Sparsity: most of the entries in $y$ are zeros.
\item $\ell_1$-minimization: minimizing $\|y\|_1$ to exploit the sparsity of $y$.
\ei

\medskip

Let us return to the flow correction problem, which is in essence equivalent to the estimation of $e_\M$.  In an extreme case, suppose all the sensors are bad, leading to large sensing errors. Without further information, it is clearly impossible to get a good estimate of $\hf$ from $f_\M$ by any means. Intuitively, however, reconstructing $\hf$ is promising if most of the sensors record consistent flow data. Mathematically speaking,  $e_\M$ is sparse. The flow correction problem thus can be viewed as sparse error correction problem \cite{CRTV_05, Cai_13}, which is similar to compressed sensing. Note that, however, the flow correction problem deviates from the traditional compressed sensing problem, where the matrix $Z$ would be random.

\section{Correction results}\label{sec:estimation}
Note that our proposed method does not take advantage of any prior information about the possible bad sensors. Apparently one can not always hope for a good estimation $f^*$ to $\hf$, even if there is only one bad sensor in the network. For instance, in the network shown in Figure \ref{fig:network}, if the sensor on link 1 gives very wrong count, then basically there is no way to reasonably correct this error because links 1 and 2 are equivalent in the topology of the network. With that said, without extra information, obtaining a good estimate of $\hf$ is possible only when the bad sensors are located at some \textit{particular} links. These locations tolerating miscount are somehow determined by the network structure. In the following, we shall introduce the concept of recoverability.

\begin{defin}
Given a network with node-link incidence matrix $A$ and monitored link set $\M$, we define the recoverability for the subset $\Ss\subseteq\M$ by
\be
{\rm Rec}(\Ss; A, \M):= \inf_{h\in\Ran(Z): \|h_{\Ss}\|_1 \not =0} \; 
{\|h_{\M\setminus\Ss}\|_1 \over \|h_{\Ss}\|_1}, \label{rec1}
\ee
which is a function of the subset $\Ss$ and also determined by both the network structure $A$ and the monitored link set $\M$. 
\end{defin}

Since $h = Z v$ for some $v\in\R^{l-n}$, then we can rewrite (\ref{rec1}) as
\be\label{rec2}
{\rm Rec}(\Ss; A, \M) = \inf_{v\in \R^{l-n}: Z_\Ss v \not = 0} {\| Z_{\M\setminus\Ss} v\|_1 \over \| Z_\Ss v\|_1 },
\ee
which resembles the classical Rayleigh quotient for the principal eigenvalue 
$\mu$ of the generalized eigenvalue problem \cite{Wein_74}: 
$Z_{\Ss}^\T\, Z_\Ss\, v = \mu \, Z_{\M\setminus\Ss}^\T\, Z_{\M\setminus\Ss}\,v$ if $\ell_2$ norm replaces the $\ell_1$ norm.  
The optimization of the ratio of two homogeneous functions of degree one has been 
studied \cite{HB_10} where an inverse power iterative algorithm was proposed. Based on \cite{HB_10}, we propose an efficient algorithm to solve problem (\ref{rec2}) which will be detailed in Appendix B.

\subsection{Exact recovery} 
We first consider the case where some sensors are bad, which introduce inconsistency of the flow data. The following Theorem \ref{thm:exact} asserts that when the bad sensors are located at certain link set $\Ss$ whose size is expected to be small, then no matter how large the errors are, we are able to exactly recover $\hf$ from $f_\M$.

\begin{thm}[Exact recovery]\label{thm:exact}
Let $\Ss:=\{i\in \M: e_i \neq 0\}$, which means miscounts only occur at the link set $\Ss$. If ${\rm Rec}(\Ss; A, \M)>1$, then the estimation $\baf$ computed by (\ref{Recon}) is equal to $\hf$.  That is, the links in $\Ss$ are robust to miscounts if ${\rm Rec}(\Ss; A, \M)>1$.
\end{thm}
The proof is omitted here, since the above theorem is a special case of Theorem \ref{thm:stable} in section \ref{sec:stable}. We remark that the lower bound for ${\rm Rec}(\Ss; A, \M)$ in the recoverability condition is sharp. Indeed the correction method can fail when ${\rm Rec}(\Ss; A, \M)=1$, as will be seen in the following example.
\begin{example}\label{eg1}
Let us consider the traffic network associated with the 3$\times$6 node-link incidence matrix
$$
A = 
\begin{bmatrix}
1 & 1 & -1 & -1&  0 & 0\\
0 & 0 & 1 & 0 & -1 & 0\\
0 & 0 & 0 & 1 & 1& -1
\end{bmatrix},
$$ 
and the ground-truth network flow $\hf = \begin{bmatrix}
300 \\
200\\
300\\
200\\
300\\
500
\end{bmatrix}$ as in Figure \ref{fig:network}, the node and links are labeled with their ID with ground truth link flows in the parentheses.

\begin{figure}
\centering
\includegraphics[trim=30mm 160mm 30mm 60mm,clip, width=0.7\linewidth]{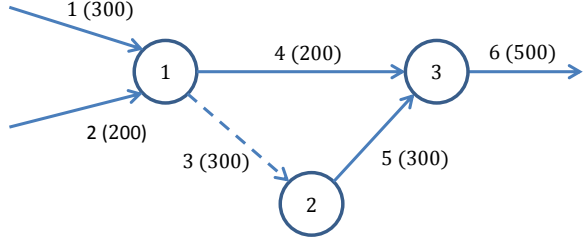}
\caption{A Toy Network. The solid links are monitored. The numbers in parentheses denote the ground-truth traffic counts.}
\label{fig:network}
\end{figure}

Then Theorem \ref{thm:ker} gives that
$$
Z = 
\begin{bmatrix}
-1 & 0 & 1 \\
1 & 0 & 0\\
0 & 1 & 0 \\
0 & -1 & 1\\
0 & 1 & 0 \\
0 & 0 & 1
\end{bmatrix}.
$$
Let the monitored link set be $\M = \{1,2,4,5,6\}$, then 
$$
Z_\M = 
\begin{bmatrix}
-1 & 0 & 1 \\
1 & 0 & 0\\
0 & -1 & 1\\
0 & 1 & 0 \\
0 & 0 & 1
\end{bmatrix}.
$$
Let the observation be $f_\M  = \begin{bmatrix}
f_1 \\
f_2\\
f_4\\
f_5\\
f_6
\end{bmatrix}
= \begin{bmatrix}
300 \\
200\\
200\\
300\\
600
\end{bmatrix}$, i.e., the observed link flow on link 6 is inflated by 100 due to sensor error. So $e_\M = f_\M - \hf_\M = \begin{bmatrix}
0 \\
0\\
0\\
0\\
100
\end{bmatrix}$, $\Ss = \{i\in \M: e_i \neq 0\} = \{6\}$, and $\M\setminus\Ss = \{1,2,4,5\}$. 

We can verify by either an analytic approach or Algorithm \ref{alg2} that the recoverability condition ${\rm Rec}(\Ss; A, \M)= 2 >1$ is satisfied. Then Theorem \ref{thm:exact} asserts that $\baf$ derived from (\ref{L1}) and (\ref{Recon}) must be equal to $\hf$. It is indeed true because $\bx=\begin{bmatrix}
200 \\
300\\
500
\end{bmatrix}$, and therefore $$\baf = Z\bx =  \begin{bmatrix}
300 \\
200\\
300\\
200\\
300\\
500
\end{bmatrix}= \hf.$$
Compare this result with the ground truth link flows, we can conclude that the errors are completely eliminated.  
\end{example}

\begin{rmk} We have two  remarks below.
\begin{itemize}
\item Without knowing the count at link 3,  i.e., $\M=\{1,2,4,5,6\}$, the proposed method would fail exact recovery if the count was corrupted at any other link except link 6. Take link 1 for example, it is easy to check that $\mathrm{Rec}(\{1\}; A, \M) = 1$. Therefore, link 1 is not guaranteed to be robust to miscount by our theory. Indeed this is the case as mentioned in the beginning of this section.

\item Suppose link 3 was also monitored, i.e., $\M= \{1, 2, \dots, 6 \}$, then any counting error at one of the links 3, 4, 5 and 6 could be accurately corrected by our method.
\end{itemize}

\end{rmk}

\subsection{Stable recovery}\label{sec:stable}
In a more realistic setting, we assume that all the elements in $e_{\M}$ are non-zeros, yet most of them are relatively small compared with the other few. This refers to approximate sparsity in compressed sensing. In this case, it is still possible for $\baf$ to be close enough to $\hf$. In another word, the estimation errors are bounded from above in this case.
\begin{thm}[Stability]\label{thm:stable}
For any $\Ss \subseteq \M$, if ${\rm Rec}(\Ss; A, \M)=\alpha>1$, then $\baf$ computed by (\ref{Recon}) obeys 
\begin{equation}\label{est}
\|\baf-\hf\|_1 \leq \lambda(\alpha, A, \M)\|e_{\M\setminus\Ss}\|_1,
\end{equation}
for some constant $\lambda(\alpha, A,\M)>0$ depending only on $\alpha$, $A$ and $\M$. Moreover, $\lambda(\alpha, A,\M)$ decreases in $\alpha$, meaning that larger recoverability leads to higher correction accuracy.
\end{thm}

In view of (\ref{est}), $f^*$ is a good estimation if $\|e_{\M\setminus\Ss}\|_1$ is small. On the other hand, the estimation error does not rely on $e_{\Ss}$. Theorem \ref{thm:exact} is essentially a corollary of Theorem \ref{thm:stable} in the special case $\|e_{\M\setminus\Ss}\|_1=0$. Therefore, it suffices to prove Theorem \ref{thm:stable} only. The proof of Theorem \ref{thm:stable} will be detailed in Appendix C, in which we derive an explicit expression for $\lambda(\alpha, A,\M)$. 

\begin{example}\label{eg2}
We consider the same setting as in Example \ref{eg1} except that the other observed data contains small sensing noise besides the large corruption at link 6. Specifically, let $f_\M = \begin{bmatrix}
302 \\
201\\
198\\
301\\
600
\end{bmatrix}$ and $e_\M = f_\M - \hf_\M = \begin{bmatrix}
302 \\
201\\
198\\
301\\
600
\end{bmatrix} - \begin{bmatrix}
300 \\
200\\
200\\
300\\
500
\end{bmatrix} = \begin{bmatrix}
2\\
1\\
-2\\
1\\
100
\end{bmatrix}$. Again we take $\Ss = \{6\}$. Since $\mathrm{Rec}(\Ss)=2>1$, it is asserted by Theorem \ref{thm:stable} that the $\ell_1$ norm of the estimation error $\|\baf-\hf\|_1$ is comparable to 
$$
\|e_{\M\setminus\Ss}\|_1 = 2+1+2+1 = 6.
$$
This is true, because $\bx = \begin{bmatrix}
201\\
303\\
503
\end{bmatrix}$ by (\ref{L1}), $\baf = \begin{bmatrix}
302\\
201\\
303\\
200\\
303\\
503
\end{bmatrix}$ , and 
$$
\|\baf-\hf\|_1 = 12.
$$
Note that the original counting error at Link 6 is 100, in sharp contrast to the error by our correction method which is just 3.
\end{example}

\section{Test Examples}\label{sec:real-world}
In this section, we provide both synthetic and real-world examples to demonstrate effectiveness of our proposed method. 

\subsection{A synthetic network}

\begin{figure}
	\centering
	\includegraphics[width=5in]{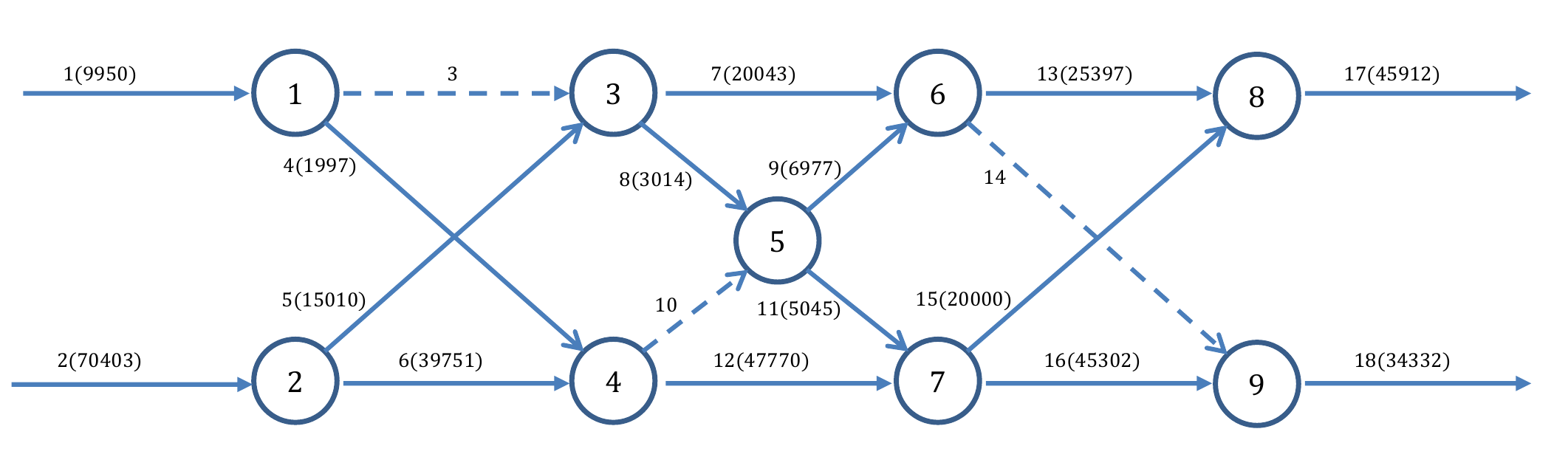}
	\caption{A parallel highway network.}
	\label{fig:para-network}
\end{figure}

 Figure \ref{fig:para-network} shows a parallel highway network \cite{Hu_09, Ng_12} with 9 nodes and 18 links among which 15 links are monitored. We create the ground-truth, observed and estimated flow data and list them in Table \ref{tb:1}.  The data on links 3, 10 and 14 are unobservable. They are marked by ``N/A'' in the table and by dashed line in the plot. The recorded data on links 6 and 16 are severely corrupted, while the other data contain small noise. So basically $\M = \{1,2,4,5,6,7,8,9,11,12,13, 15,16,17,18\}$ and $\Ss = \{6,16\}$.  It is clear that our estimation by Algorithm \ref{alg} is fairly close to the ground-truth, and the miscounts on links 6 and 16 are successfully detected. In fact, we can check by Algorithm \ref{alg2} that the recoverability condition $\mathrm{Rec}(\Ss; A, \M)= 1.5>1$ holds. Therefore, Theorem \ref{thm:stable} provides guarantee for our correction result.

\begin{table}[]
	\centering
	\caption{Computational results for Example 1. Links with corrupted data are labeled with *.}
	\label{tb:1}
	\begin{tabular}{|c|c|c|c|c|}
		\hline
		Link ID & Ground-truth & Observation & Estimation & Estimation Error  \\ \hline
		1       &    10000   & 9950      & 9950     & -10               \\ \hline
		2       &    70000 &   69887      & 69887    & -113            \\ \hline
		3       &    8000   &  N/A        &  7953    & -47            \\ \hline
		4       &    2000   & 1997      & 1997      & -3              \\ \hline
		5       &    15000    &15010    & 15104     & 104          \\ \hline
		6*       &    55000   &  {\bf 39751}      & 54783      & -217      \\ \hline
		7       &    20000     &   20043   & 20043     & 43         \\ \hline
		8       &    3000      & 3014    & 3014      & 14           \\ \hline
		9       &    7000       &  6977    & 6977     & 23           \\ \hline
		10      &    9000     & N/A      & 9009      & 9             \\ \hline
		11      &    5000     & 5045     & 5046    & 46              \\ \hline
		12      &    48000   &  47770    & 47771      & -229       \\ \hline
		13      &    25500    & 25397     & 25505    & 5           \\ \hline
		14      &     1500      & N/A         & 1515     & 15          \\ \hline
		15      &   20000   &  20000      & 20000     & 0          \\ \hline
		16*      &    33000  & {\bf 45302}       & 32817      & -183       \\ \hline
		17      & 45500    & 45912     & 45505    & 5             \\ \hline
		18      &   34500  & 34332      & 34332      & -168          \\ \hline
	\end{tabular}
\end{table}

\subsection{A real-world example}

\begin{figure}
	\centering
	\includegraphics[width=5in]{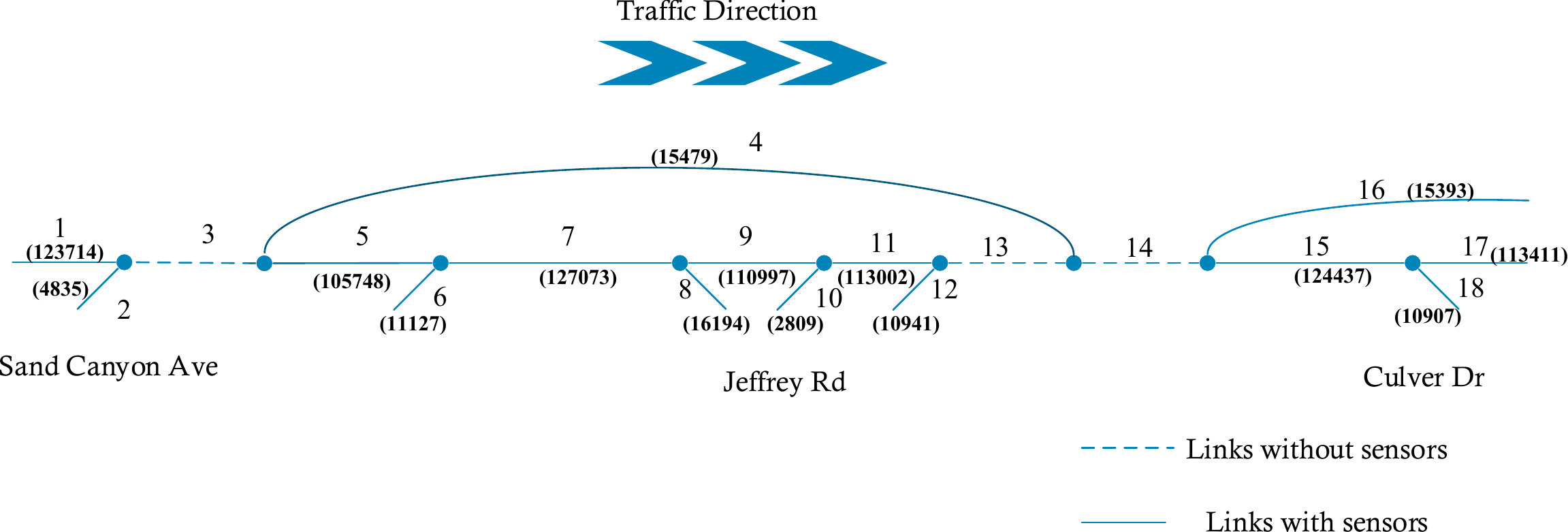}
	\caption{A road network on I-405 northbound in the city of Irvine.}
	\label{fig:405-network}
\end{figure}

  The daily cumulative flow data in this example is from Caltrans Performance Measurement System (PeMS) database, collected on I-405 northbound in the city of Irvine, on April 28, 2016. The network has 18 links and 9 nodes as illustrated in Figure \ref{fig:405-network}. The loop detectors are installed on all links except for links 3, 13, and 14, which are represented by dashed lines. The links are labeled with their IDs and corresponding observed flows in the parentheses. 

The estimated link flows by (\ref{L1}) and (\ref{Recon}) are compared with the observed link flows in Table \ref{tb:estimation}, where unobserved links flows are marked by ``N/A''. Our correction result shows that the estimation error at link 6 is much larger than all other links. Since there is no ground-truth data available in this example, we can not check the correction quality directly. However, link 6 is flagged as unhealthy sensor by PeMS, which is consistent with our estimation. On the other hand, if link 6 is indeed the only unhealthy sensor, the quality of the estimated link flow listed in Table \ref{tb:estimation} is guaranteed by Theorem \ref{thm:stable}, in which we have $\M = \{1,2,4,5,6,7,8,9,10,11,12,15,16,17,18\}$ and $\Ss = \{6\}$. It can be verified that the recoverability condition ${\rm Rec}(\Ss; A, \M)=2 > 1$ holds. 

\begin{table}[]
	\centering
	\caption{Computational results for Example 2. }
	\label{tb:estimation}
	\begin{tabular}{|c|c|c|c|c|}
		\hline
		Link ID & Observation & Estimation & Difference & Percentage Difference\\ \hline
		1       & 123714      & 123714     & 0          & 0.0\%      \\ \hline
		2       & 4835        & 4835       & 0          & 0.0\%      \\ \hline
		3       & N/A         & 128549     & N/A        & N/A        \\ \hline
		4       & 15479       & 15479      & 0          & 0.0\%      \\ \hline
		5       & 105748      & 113070     & 7322       & 6.9\%      \\ \hline
		6       & 11127       & 13661      & 2534       & 22.8\%     \\ \hline
		7       & 127073      & 126731     & -342       & -0.3\%     \\ \hline
		8       & 16194       & 16194      & 0          & 0.0\%      \\ \hline
		9       & 110997      & 110537     & -460       & -0.4\%     \\ \hline
		10      & 2809        & 2757       & -52        & -1.9\%     \\ \hline
		11      & 113002      & 113295     & 293        & 0.3\%      \\ \hline
		12      & 10941       & 10941      & 0          & 0.0\%      \\ \hline
		13      & N/A         & 124236     & N/A        & N/A        \\ \hline
		14      & N/A         & 139715     & N/A        & N/A        \\ \hline
		15      & 124437      & 124322     & -115       & -0.1\%     \\ \hline
		16      & 15393       & 15393      & 0          & 0.0\%      \\ \hline
		17      & 113411      & 113413     & 2          & 0.0\%      \\ \hline
		18      & 10907       & 10909      & 2          & 0.0\%      \\ \hline
	\end{tabular}
\end{table}

\section{Conclusion}\label{sec:conclusion}

In this study, we systematically studied the link flow correction problem in 
a traffic network based on flow conservation. The problem is formulated as an  $\ell_1$-minimization problem, 
in which the differences between the estimated and observed link flows are minimized. 
We introduced the recoverability concept for a subset of links and  specifically derived the recoverability condition for exactly retrieving the missing data: 
when certain sensors are malfunctioning, no matter how large the errors are, 
the ground truth flow can be exactly recovered. That is, some links are robust to miscounts. 
Furthermore, when small errors are present in observed link flows, 
the estimation error bound is found such that we can estimate the link flows that 
are close enough to ground-truth under the recoverability condition. We also showed an efficient 
algorithm for computing recoverability.

A few follow-up study topics can be interesting both theoretically and practically. In addition to the $\ell_1$ norm, it will be interesting to investigate the feasibility and efficiency of  other sparsity promoting penalty functions for formulating and solving the flow correction problem. The recoverability defined in (\ref{rec1}) is central to the flow correction problem, as it determines whether exact recovery is possible or not (see Theorem \ref{thm:exact}) and also the error bound in stable recovery (see (\ref{est})). In the future we will be interested in examining with Algorithm 2 how the road network's structure impacts the recoverability of a subset of links, and such a study could provide guidelines for installing flow counting sensors especially in a large-scale network.

\section*{Acknowledgments}
Yin and Xin were partially supported by NSF grants DMS-1522383 and IIS-1632935. 
Yin was also supported by ONR grant N000141617157. 
We would like to thank the referees for their constructive comments.

\section*{Appendix A. ADMM for solving (\ref{L1})}
The following alternating direction method of multipliers (ADMM) \cite{Boyd_11} is an thresholding-based iterative algorithm. In Algorithm \ref{alg}, $z\in\Rm$ and $u\in\Rm$ are auxiliary variables. 'shrink' is the so-called soft-thresholding operator on $\Rm$. For any $z\in\Rm$ and $r>0$, $\mathrm{shrink}(z,r)$ performs component-wise operation on $z$ given by
$$
\left(\mathrm{shrink}(z,r)\right)_i = \mathrm{sign}(z_i)\max\{|z_i|-r,\; 0\}, \quad i = 1,\dots, m.
$$
The algorithm stops after some maximum number of iterations.
\begin{algorithm}
\caption{ADMM for solving $\min_{x\in\R^{l-n}} \|Z_{\M}x - f_{\M}\|_1$.}
\label{alg}
Input: $Z_{\M}, \; f_\M, \; \delta > 0$\\
Initialize: $x^{(0)}, \; z^{(0)}, \; u^{(0)}$
\begin{algorithmic}
\FOR {$i = 0,1,\dots, k_1-1$}
\STATE $x^{(i+1)} = (Z_{\M}^\T Z_{\M})^{-1}Z_{\M}^\T(f_{\M} + z^{(i)} - u^{(i)})$
\STATE $z^{(i+1)} = \mathrm{shrink}(Z_\M x^{(i+1)} - f_\M + u^{(i)}, \frac{1}{\delta})$
\STATE $u^{(i+1)} = u^{(i)} + Z_\M x^{(i+1)} - z^{(i+1)} -f_\M$
\ENDFOR
\end{algorithmic}
Output: $x^* = x^{(k_1)}$
\end{algorithm}

\section*{Appendix B. An inverse power algorithm for solving (\ref{rec2})}
We present Algorithm \ref{alg2} to solve the following optimization problem (\ref{rec2}):
$$
\min_{v\in\R^{l-n}} \; \frac{\|Z_{\M\setminus\Ss} v\|_1}{\|Z_{\Ss} v\|_1 }.
$$
\begin{algorithm}
\caption{An inverse power algorithm \cite{HB_10} for solving (\ref{rec2}).}
\label{alg2}
Input: $Z_{\M\setminus\Ss}, \; Z_{\Ss}$\\
Initialize: $v^{(0)}, \; \lambda^{(0)} = \frac{\|Z_{\M\setminus\Ss} v^{(0)}\|_1}{\|Z_{\Ss} v^{(0)}\|_1}$
 
\begin{algorithmic}
\FOR {$i = 0,1,\dots, k_2-1$}
\STATE $v^{(i+1)} = \arg\min_v \; \|Z_{\M\setminus\Ss} v\|_1 - \lambda^{(i)}\langle Z_{\Ss}^\T\mathrm{sign}(Z_{\Ss}v^{(i)}) , v\rangle \quad \mbox{subject to} \quad \|v\|\leq1$\\
\STATE $\lambda^{(i+1)} =\frac{\|Z_{\M\setminus\Ss} v^{(i+1)}\|_1}{\|Z_{\Ss} v^{(i+1)}\|_1}$
\ENDFOR
\end{algorithmic}
Output: $\lambda^* = \lambda^{(k_2)}$
\end{algorithm}
The output $\lambda^*$ is the optimal objective value in (\ref{rec2}), i.e., $\mathrm{Rec}(\Ss; A, \M)$. 
Note that in Algorithm \ref{alg2}, updating $v$ under the unit ball constraint is non-trivial and requires extra effort. 
We write an ADMM solver for this subproblem in Algorithm \ref{alg3} below.

\begin{algorithm}
\caption{ADMM for updating $v$.}
\label{alg3}
Input: $Z_{\M\setminus\Ss}, \; Z_{\Ss}, \; b = \lambda^{(i)} Z_{\Ss}^\T\mathrm{sign}(Z_{\Ss}v^{(i)}) $ from Algorithm \ref{alg2}, and $ \delta > 0$\\
Initialize: $v^{(0)}, \; z^{(0)}, \; u^{(0)}$
\begin{algorithmic}
\FOR {$j = 0,1,\dots, k_3-1$}
\STATE $v^{(j+1)} = (Z_{\M\setminus\Ss}^\T Z_{\M\setminus\Ss})^{-1}\left(Z_{\M\setminus\Ss}^\T(z^{(j)} + \frac{u^{(j)}}{\delta}) + \frac{b}{\delta}\right)$
\STATE $v^{(j+1)} = \frac{v^{(j+1)}}{\|v^{(j+1)}\|}$ \quad \textbf{if} $\|v^{(j+1)}\|>1$
\STATE $z^{(j+1)} = \mathrm{shrink}(Z_{\M\setminus\Ss} v^{(j+1)}  + \frac{u^{(j)}}{\delta}, \frac{1}{\delta})$
\STATE $u^{(j+1)} = u^{(j)} + \delta(z^{(j+1)}-Z_{\M\setminus\Ss}v^{(j+1)}) $
\ENDFOR
\end{algorithmic}
Output: $v^{(i+1)}$ in Algorithm \ref{alg2}
\end{algorithm}

\section*{Appendix C. Technical proofs}
\begin{proof}[{\bf Proof of Theorem \ref{thm:ker}}]
To prove $Z = 
\begin{bmatrix}
I_{(l-n)} \\
-(A^{\K^c})^{-1}A^{\K}
\end{bmatrix}
\in\R^{l\times(l-n)}$ gives a basis of $\Ker(A)$, it suffices to show that
\bi
\item[1.] $AZ = O$ is a zero matrix. It is true since $AZ = [A^{\K}, \; A^{\K^c}] \begin{bmatrix}
I_{(l-n)} \\
-(A^{\K^c})^{-1}A^{\K}
\end{bmatrix} = A^{\K} - A^{\K^c}(A^{\K^c})^{-1}A^{\K} = O$.
\item[2.] $Z$ has full rank, i.e., $\rank(Z) = l-n$. This is also true because, on one hand $\rank(Z)\leq l-n$, on the other hand, $\rank(Z)\geq \rank(I_{(l-n)}) = l-n$ since $I_{(l-n)}$ is a submatrix of $Z$.
\ei
Then by (\ref{Recon}), we have 
$$
f^* = \begin{bmatrix}
I_{(l-n)} \\
-(A^{\K^c})^{-1}A^{\K}
\end{bmatrix}x^* = \begin{bmatrix}
x^* \\
-(A^{\K^c})^{-1}A^{\K}x^*
\end{bmatrix}.
$$
Since $\hf = \begin{bmatrix}
\hf^{\K} \\
\hf^{\K^c}
\end{bmatrix}$, we conclude that $x^*$ is an estimate of $\hf^{\K}$.
\end{proof}

\begin{proof}[{\bf Proof of Theorem \ref{thm:stable}}]
Suppose $f^* = \hf + v$, since
$f^*, \;\hf\in\Ran(Z)$, then we have 
$$
v\in\Ran(Z), \; v_\M\in\Ran(Z_\M)
$$
Moreover, since $f_\M^* = Z_\M x^*$ and $\hf_\M\in\Ran(Z_\M)$, (\ref{L1}) implies that
\begin{equation}\label{ineq1}
\|f_\M^*-f_\M\|_1 = \|Z_\M x^* -f_\M\|_1 \leq \|\hf_\M - f_\M\|_1.
\end{equation}
Keep in mind that $e_\M = f_\M - \hf_\M$ is the sensing error, so on the right hand side of (\ref{ineq1}),
\begin{equation}\label{ineq2}
\|\hf_\M-f_\M\|_1 = \|e_\M\|_1 = \|e_\Ss\|_1 + \|e_{\M\setminus\Ss}\|_1,
\end{equation}
and on the left hand side,
\begin{align}
\|f_\M^*-f_\M\|_1 &  = \|(\hf + v)_\M - f_\M\|_1 = \|v_\M - e_\M\|_1 \notag \\
& = \|v_\Ss - e_\Ss\|_1 + \|v_{\M\setminus\Ss} - e_{\M\setminus\Ss}\|_1 \notag\\
& \geq \|e_\Ss\|_1 - \|v_\Ss\|_1 + \|v_{\M\setminus\Ss}\|_1 -\|e_{\M\setminus\Ss}\|_1 \label{ineq3}
\end{align}
In (\ref{ineq3}), we used the triangle inequality for $\ell_1$ norm. Combining (\ref{ineq1}), (\ref{ineq2}), and (\ref{ineq3}), we have
$$
\|e_\Ss\|_1 + \|e_{\M\setminus\Ss}\|_1\geq \|e_\Ss\|_1 - \|v_\Ss\|_1 + \|v_{\M\setminus\Ss}\|_1 -\|e_{\M\setminus\Ss}\|_1
$$
or 
\begin{equation}\label{ineq4}
2\|e_{\M\setminus\Ss}\|_1 \geq - \|v_\Ss\|_1 + \|v_{\M\setminus\Ss}\|_1
\end{equation}
By the assumption that ${\rm Rec}(\Ss; A, \M)=\alpha >1$, we have $\alpha\|h_{\Ss}\|_1 \leq \|h_{\M\setminus\Ss}\|_1$ holds for all $h\in\Ran(Z_\M)$. Since $v_\M\in\Ran(Z_\M)$ as aforementioned, we have $\|v_\M\|_1\geq (1+\alpha)\|v_\Ss\|_1$, then it follows from (\ref{ineq4}) that
$$
2\|e_{\M\setminus\Ss}\|_1 \geq \|v_\M\|_1 - 2\|v_\Ss\|_1\geq (1-\frac{2}{1+\alpha})\|v_\M\|_1,
$$
and thus
\begin{equation}\label{ineq5}
\|(f^*-\hf)_\M\|_1 = \|v_\M\|_1\leq \frac{2(\alpha+1)}{\alpha-1}\|e_{\M\setminus\Ss}\|_1.
\end{equation}

In what follows, we derive an upper bound for $\|(f^*-\hf)_{\M^c}\|_1$. Without loss generality, suppose $A = [A^{\K}, \; A^{\K^c}]$ with $\K\subseteq\M$ being any base set. Since \textbf{b}\textbf{o\textbf{\textbf{\textbf{}}}}th $f^*$ and $\hf$ obey flow conservation, we have
$$
\0_{(n)} = A(f^* - \hf) = [A^{\K}, \; A^{\K^c}]\begin{bmatrix}
(f^* - \hf)_\K \\
(f^* - \hf)_{\K^c} 
\end{bmatrix},
$$ 
which gives
$$
(f^*-\hf)_{\K^c} = -(A^{\K^c})^{-1}A^{\K}(f^*-\hf)_{\K}.
$$
Since $\K\subseteq\M$, we have $\M^c\subseteq\K^c$. Therefore, $(f^*-\hf)_\K$ is contained in $(f^*-\hf)_\M$, and $(f^*-\hf)_{\M^c}$ is contained in $(f^*-\hf)_{\K^c}$.
Using the above facts, we have
\begin{align}\label{ineq6}
\|(f^*-\hf)_{\M^c}\|_1& \leq \|(f^*-\hf)_{\K^c}\|_1
= \|-(A^{\K^c})^{-1}A^{\K}(f^*-\hf)_{\K}\|_1 \notag\\
& \leq \|(A^{\K^c})^{-1}A^{\K}\|_1\|(f^*-\hf)_\K\|_1 \leq \|(A^{\K^c})^{-1}A^{\K}\|_1\|(f^*-\hf)_\M\|_1 \notag\\
& \leq \frac{2(\alpha+1)}{\alpha-1}\|(A^{\K^c})^{-1}A^{\K}\|_1 \|e_{\M\setminus\Ss}\|_1.
\end{align}
In the second inequality above, $\|(A^{\K^c})^{-1}A^{\K}\|_1$ is the operator norm of $(A^{\K^c})^{-1}A^{\K}$ induced by $\ell_1$ norm. And in the last inequality, we used (\ref{ineq5}).

Finally, combining (\ref{ineq5}) and (\ref{ineq6}) gives that
$$
\|f^*-\hf\|_1 \leq \frac{2(\alpha+1)}{\alpha-1}(\|(A^{\K^c})^{-1}A^{\K}\|_1 + 1)\|e_{\M\setminus\Ss}\|_1.
$$
Note that the above inequality holds for all base set $\K\subseteq\M$. Therefore,
$$
\|f^*-\hf\|_1 \leq \frac{2(\alpha+1)}{\alpha-1}\min_{\overset{\K\subseteq\M}{\K \mbox{ is base set}}}\{\|(A^{\K^c})^{-1}A^{\K}\|_1 + 1\}\|e_{\M\setminus\Ss}\|_1,
$$
which concludes the proof.
\end{proof}

\end{document}